\documentclass[twocolumn,amsmath,amssymb,footinbib,superscriptaddress]{revtex4-1}

\usepackage{graphicx}% Include figure files
\usepackage{dcolumn}% Align table columns on decimal point
\usepackage{bm}% bold math
\usepackage{color}
\usepackage{epstopdf}
\usepackage{amsmath}
\usepackage{lipsum}
\usepackage[colorlinks=true,citecolor=blue,urlcolor=blue,linkcolor=blue]{hyperref}

\begin{document}

\title{Intrinsic coherent acoustic phonons in indirect band gap semiconductors Si and GaP}

\author{Kunie Ishioka}
\email{ishioka.kunie@nims.go.jp}
\affiliation{National Institute for Materials Science, Tsukuba, 305-0047 Japan}

\author{Avinash Rustagi}
\affiliation{Department of Physics, University of Florida, Gainesville, FL 32611 USA}

\author{Ulrich H{\"o}fer}
\affiliation{Faculty of Physics and Materials Sciences Center, Philipps-Universit{\"a}t Marburg, 35032 Marburg, Germany}

\author{Hrvoje Petek}
\affiliation{Department of Physics and Astronomy, University of Pittsburgh, Pittsburgh, PA 15260 USA}

\author{Christopher J. Stanton}
\affiliation{Department of Physics, University of Florida, Gainesville, FL 32611 USA}

\date{\today}

\begin{abstract}

We report on the intrinsic optical generation and detection of coherent acoustic phonons at (001)-oriented bulk Si and GaP without metallic phonon transducer structures.  Photoexcitation by a 3.1-eV laser pulse generates a normal strain pulse within the $\sim$100-nm penetration depth in both semiconductors.  The subsequent propagation of the strain pulse into the bulk is detected with a delayed optical probe as a periodic modulation of the optical reflectivity.  Our theoretical model explains quantitatively the generation of the acoustic pulse via the deformation potential electron-phonon coupling and detection in terms of the spatially and temporally dependent photoelastic effect  for both semiconductors.  
Comparison with our theoretical model reveals that the experimental strain pulses have finite build-up times of 1.2 and 0.4 ps for GaP and Si, which are comparable with the time required for the photoexcited electrons to transfer to the lowest X valley through intervalley scattering.  
The deformation potential coupling related to the acoustic pulse generation for GaP is estimated to be twice as strong as that for Si from our experiments, in agreement with a previous theoretical prediction.  

\end{abstract}

\maketitle

\section{INTRODUCTION}

Photoexcitation of a solid with a femtosecond laser pulse creates carriers in a non-equilibrium distribution, which transfer energy to the lattice and thereby excite phonons on time scales ranging from femto- to nanoseconds.  One of the intriguing ultrafast electron-phonon interactions triggered by the photoexcitation is the generation of coherent phonons.   Coherent optical phonons are non-propagating lattice oscillations
with wavevectors $q\sim0$ and discrete THz frequencies \cite{Foerst}.
Coherent acoustic phonons, by contrast, are ultrasonic strain pulses that can have a broad spectrum from GHz up to a few THz frequencies and wide range of wavevectors $q$.
Propagation of the coherent acoustic phonons has been studied in various materials by means of time-resolved optical \cite{Thomsen86,Mante15,Ruello15,Matsuda15} and x-ray diffraction \cite{Reis,Rpetruck99} measurements.  Such studies are motivated by optical determination of the mechanical properties of solids \cite{Ogi07,Hudert09}, optical control of acoustic waves in solids \cite{Nelson82}, acoustic tomography of buried interfaces and objects hidden under surfaces \cite{Wright92,Yahng02}, and ultrafast optical control of the  piezoelectric effect \cite{Sun01}.  

Coherent acoustic phonons can be excited efficiently at a metal surface, where a laser pulse is absorbed within its short optical penetration depth, typically $\lesssim$10 nm, and thereby heats up the lattice to sudden expansion (thermal stress) \cite{Thomsen86}.  This thermoelastic effect, however, is generally much weaker in semiconductors, so most of the previous studies have resorted to  putting metallic thin films on top of semiconductors to act as opto-acoustic transducers.  Such scheme enables the generation of intense acoustic pulses and thereby allows a detailed study of acoustic diffraction and dispersion in semiconductors \cite{Daly04,Bosco02,Devos04}.
One can alternatively use semiconductor heterostructures, such as InGaN/GaN quantum wells   and GaN \textit{p-n} junctions \cite{Sun00,Liu05,Wen09,*Wen11,Chen12,Kim12}, as transducers; in these cases intense acoustic phonons are generated via the selective absorption of the pump pulse in the quantum wells and the screening of the built-in piezoelectric field. 

The intrinsic optical generation of coherent acoustic phonons in simple bulk semiconductors, by contrast, has been hardly explored by conventional optical measurements.  The exceptions are the direct band gap GaAs \cite{Wright01,Matsuda04,Babilotte10,Young12,Vaudel14} and GaN \cite{Wu06,*Wu07}, which were excited with ultraviolet pulses for photon energies above their band gaps.
In both cases, the penetration depths of the pump lights was about as short as those in metals, and the steep distribution of the photocarriers along the depth direction lead to the efficient generation the coherent acoustic phonons.  The generation mechanism was attributed mainly to a deformation potential interaction of photoexcited carriers with the lattice based on theoretical modeling for both GaAs and GaN \cite{Wright01,Wu06,*Wu07}. 

In principle, It should also  be possible to excite acoustic phonons in \emph{indirect} band gap semiconductors such as Si, whose deformation potential coupling constant is comparable to that of GaAs \cite{Cardona87}.  Until now there has been no report of such pulse generation and detection in Si without introduction of a metallic transducer structure, however. 
A previous optical deflection study reported only a gradual surface contraction on a nanosecond time scale following photoexcitation of bare Si surface, but coherent acoustic phonons were not observed \cite{Wright95}.  

In the present study, we investigate the intrinsic coherent acoustic phonon generation and detection in bulk indirect band gap semiconductors Si and GaP using a pump-probe reflectivity detection technique.  
We generate coherent acoustic phonons in GaP and Si by photoexciting their surfaces with a 3.1-eV pump pulse, and monitor their propagation into the bulk through modulation of the reflected intensity of the probe pulse.  Our theoretical model quantitatively reproduces the experimentally observed reflectivity responses for both GaP and Si, and offers insight into the deformation potential coupling that contributes to the acoustic phonon generation.

\section{EXPERIMENTAL METHODS}

\begin{figure}
\includegraphics[width=0.375\textwidth]{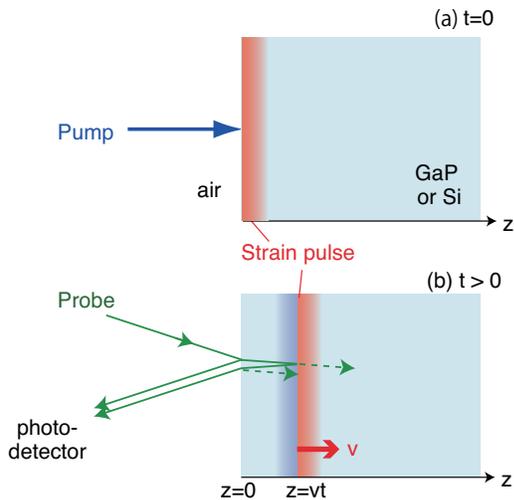}
\caption{\label{Scheme} (Color Online.) Schematic of the experiment.  (a) Pump light   generates an acoustic pulse at the surface ($z$=0) at $t$=0.  (b) The acoustic pulse propagates into the bulk at the speed of sound $v$.  The probe light is reflected by the acoustic pulse, which is positioned at $z=vt$ for $t>0$, as well as by the surface.  
The incident angle of the probe light in the lower panel is exaggerated for clarity.}
\end{figure}

\begin{figure}
\includegraphics[width=0.45\textwidth]{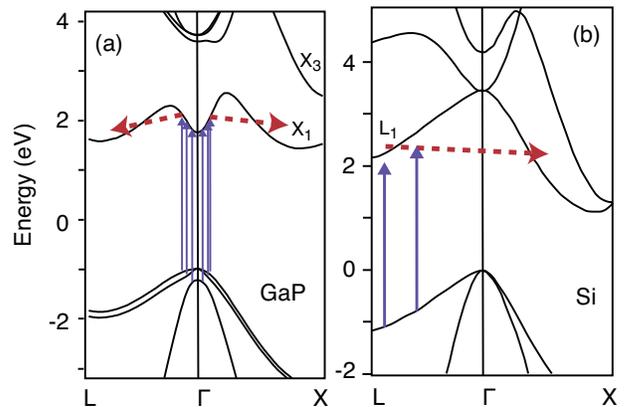} 
\caption{\label{band}  (Color Online.) Schematic band structures of GaP (a) and Si (b) near the fundamental band gaps \cite{Ishioka15,Sieh85}.  Solid and broken arrows indicate the major transitions with 3.1-eV photons and the relaxation pathways discussed in the text.}
\end{figure}

The samples studied are (001)-oriented $n$-doped Si and GaP single crystal wafers.
Pump-probe reflectivity measurements are performed in a near back-reflection configuration under ambient conditions, as schematically shown in Fig.~\ref{Scheme}.  
For the one-color measurements, the second harmonic of a Ti:sapphire oscillator with 3.1-eV photon energy (400-nm wavelength), 10-fs duration and 80 MHz repetition rate is used for both pump and probe pulses.  The 3.1-eV photons can excite carriers across the direct band gap at the $\Gamma$ point for GaP, as shown in Fig.~\ref{band}a.  For Si, the pump photon energy is slightly less than the $E_0'$ and $E_1$ gaps at the $\Gamma$ and $L$ points \cite{Adachi88}, but can excite carriers along the $L$ valleys assisted by small-$q$ phonons, as shown in Fig.~\ref{band}b.  The optical penetration depths for the 3.1-eV light in GaP and Si are $\alpha_\textrm{GaP}^{-1}$=116 nm and $\alpha_\textrm{Si}^{-1}$=82 nm \cite{Aspnes83}.  The pump and probe laser spots on the sample are $\sim23 \mu$m in diameter.  Pump-induced change in the reflectivity $\Delta R$ is measured as a function of time delay between pump and probe pulses using a fast scan technique.  This scheme allows us to monitor the ultrasonic pulses in the first few tens of ps with 10 fs time resolution \cite{Ishioka08}.   

For the two-color measurements, the second harmonic of a regenerative amplifier with 3.1-eV photon energy, 150-fs duration and 100 kHz repetition rate is used as the pump pulse, whereas the output of an optical parametric amplifier with tunable wavelength in the visible range serves as the probe.  The pump and probe laser spots on the sample are $\sim$250 and $\sim180 \mu$m in diameter.  $\Delta R$ is measured as a function of time delay between pump and probe pulses using a slow scan technique.  Since the visible probe light penetrates much deeper than the 3.1-eV pump, this scheme allows us to monitor the acoustic pulses up to the sub-ns time scales.

\section{Experimental RESULTS}\label{exp}
\subsection{One-color pump-probe measurements}

\begin{table}
\caption{\label{Const} Material parameters for GaP and Si used in the present study.}
\begin{ruledtabular}
\begin{tabular}{ccccll}
&GaP&Si&unit&&Ref.\\
\hline
$\alpha$&8.628 &12.162&$\times10^4$cm$^{-1}$&at 3.1 eV&\cite{Aspnes83}\\
$n$&4.196&5.570&& at 3.1 eV&\cite{Aspnes83}\\
$v$&5.847&8.4332&$\times10^5$cm/s &&\cite{Weil68,McSkimin64}\\
$E_g$&2.26&1.12&eV& at 300 K&\cite{Adachi87,Adachi88}\\
$\partial E_g/\partial p$&11&5&$\times10^{-11}$eV/Pa&&\cite{Zallen67}\\
$\beta$&4.65&2.6&$\times10^{-6}$K$^{-1}$&&\cite{Deus83,Okada84}\\
$C$&0.43&0.7&J g$^{-1}$K$^{-1}$&&\cite{Flubacher59}\\
\end{tabular}
\end{ruledtabular}
\end{table}

\begin{figure}
\includegraphics[width=0.48\textwidth]{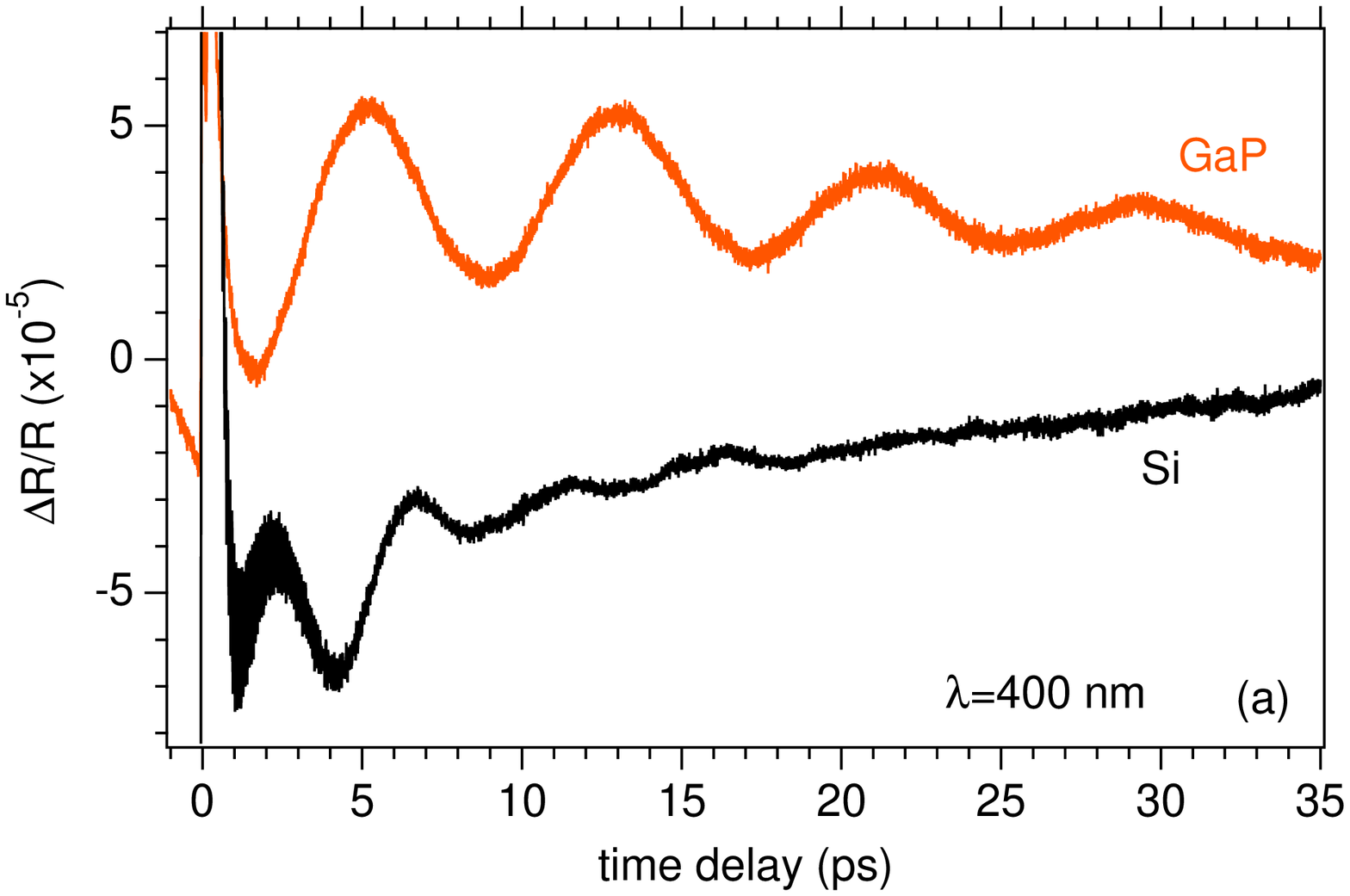}
\includegraphics[width=0.475\textwidth]{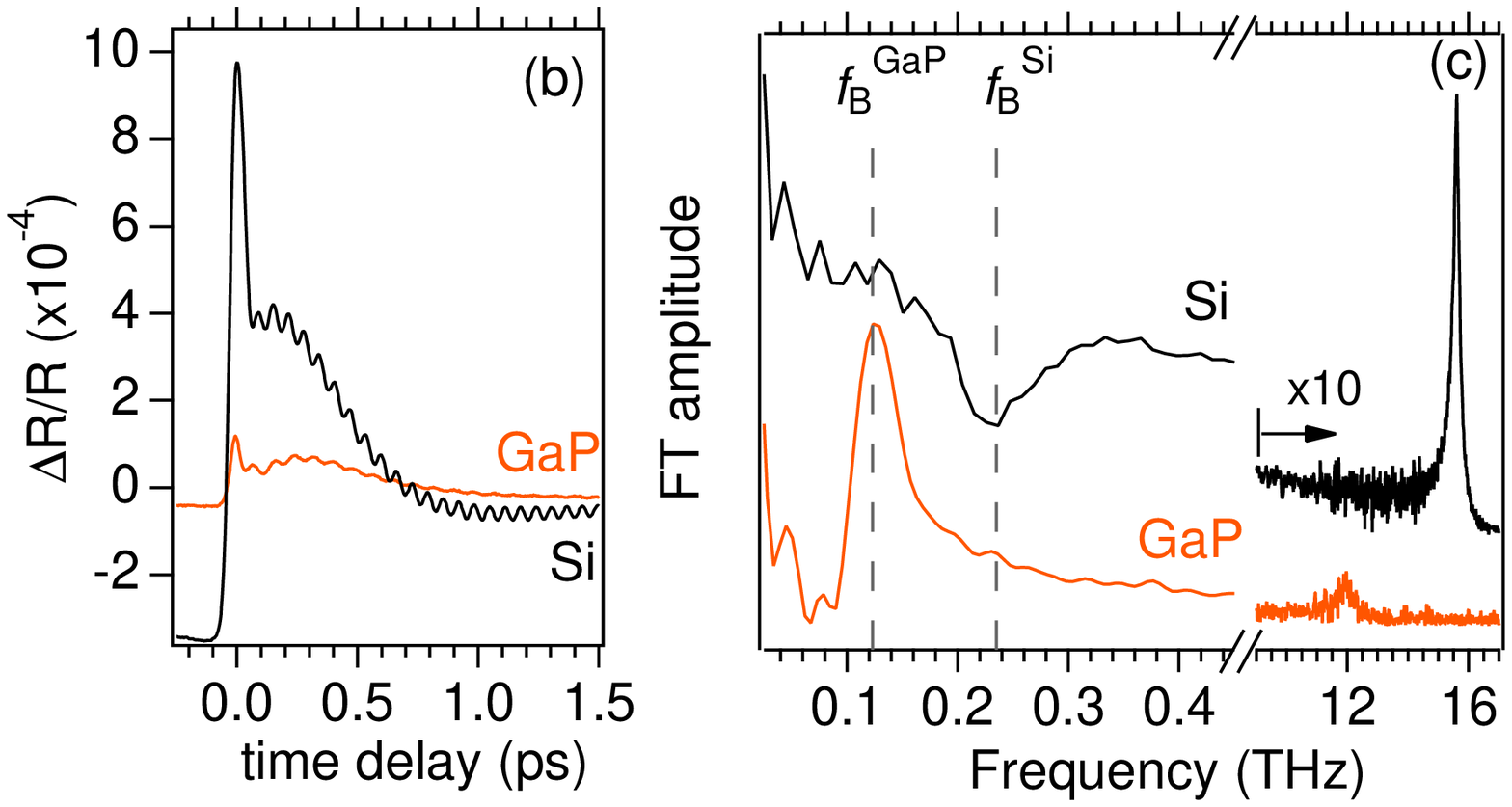}
\caption{\label{TD400} (Color Online.) (a,b) Reflectivity changes of (001)-oriented GaP and Si  pumped and probed at 3.1 eV (400 nm). Pump densities are 30 and 90 $\mu$J/cm$^2$ for GaP and Si.  (a) and (b) show identical traces with different vertical and horizontal scales.  (c) Fourier-transformed spectra of the reflectivity changes after $t$=0.1 ps.  Broken lines indicate the frequencies $f_B^\textrm{GaP}$ and $f_B^\textrm{Si}$ given by eq.(\ref{freq}) for GaP and Si. }
\end{figure}

We first report one-color pump-probe measurements with 3.1-eV pulses that monitor the coherent phonon dynamics in  GaP and Si for the first few tens of picoseconds, which are shown in Fig.~\ref{TD400}.  In the present study we  photoexcite GaP at a lower pump density than Si, because the GaP surface is more easily damaged by the laser irradiation.  For the first picosecond [Fig.~\ref{TD400}b], the reflectivity traces exhibit a non-oscillatory electronic response and, on top of that, a fast periodic modulation due to the generation of coherent longitudinal optical (LO) phonons at 12 and 15.6 THz for GaP and Si, which have been previously reported elsewhere \cite{Hase03,Ishioka15}.  

On a longer time scale [Fig.~\ref{TD400}a], the reflectivity traces clearly show much slower ($<$1 THz) periodic modulations, whose amplitudes ($\Delta R/R\sim10^{-5}$) are comparable or larger than those of the coherent LO phonons.  We attribute the slow modulations to the interference between the probe reflections from the front surface of the sample and a propagating acoustic pulse, also known as the Brillouin oscillation, which is schematically explained  in Fig.~\ref{Scheme}.  The frequency $f_B$ of such interference is given, in the case of the normal incidence, by \cite{Thomsen86}:
\begin{equation}\label{freq}
f_B =\dfrac{2nv}{\lambda}, 
\end{equation}
where $n$ is the refractive index, $v$, the longitudinal acoustic (LA) phonon velocity, and $\lambda$, the probe wavelength in air.  
With $n$ and $v$ listed in Table~\ref{Const}, 
we obtain $f_B^\textrm{GaP}$=123 GHz and $f_B^\textrm{Si}$=235 GHz.  Fourier-transformed (FT) spectra in Fig.~\ref{TD400}c indicate good agreement between the experimental and calculated frequencies, confirming the origin of the modulations.  We note that the FT spectrum for Si exhibits a dip, instead of a peak, at $f_B^\textrm{Si}$.  This is because of interference between the periodic oscillation and the large non-oscillatory electronic response appearing in the first picosecond.

\subsection{Two-color pump-probe measurements}

\begin{figure}
\includegraphics[width=0.47\textwidth]{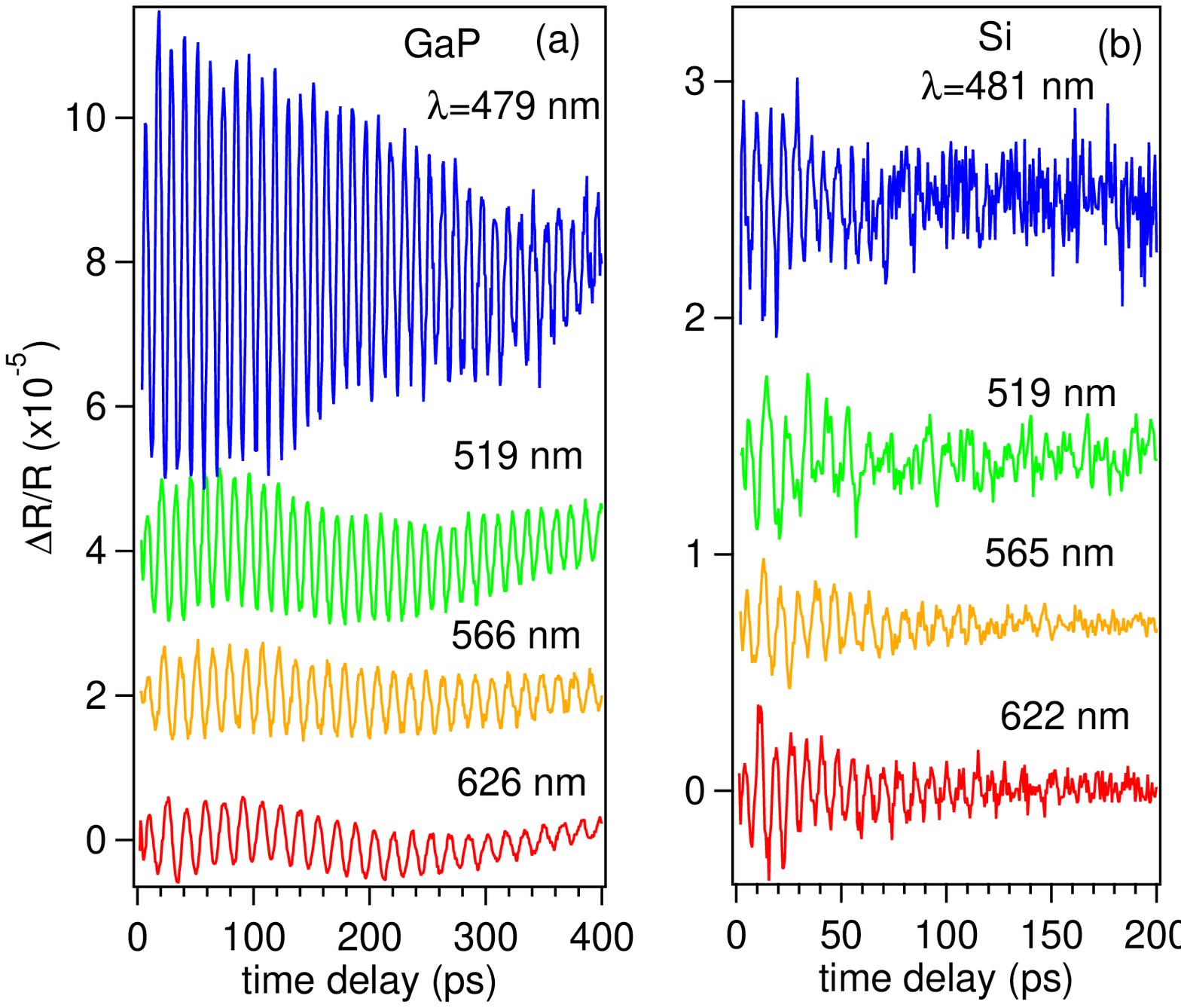}\\
\vspace{4mm}
\includegraphics[width=0.4\textwidth]{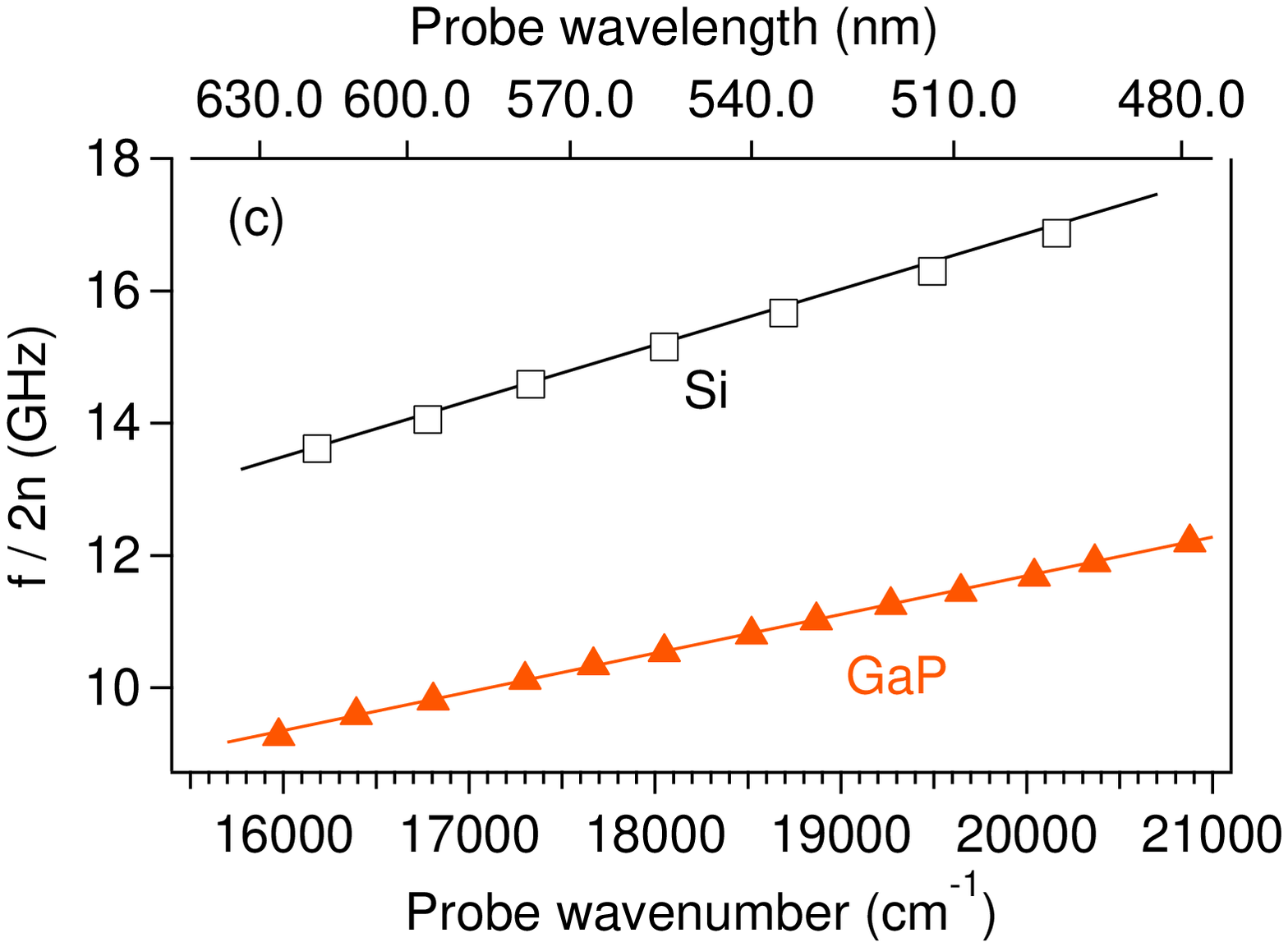}
\caption{\label{TDVis} (Color Online.) (a,b) Oscillatory parts of the reflectivity changes pumped at 3.1 eV and probed at different wavelength $\lambda$; for GaP (a) and Si (b).  Pump densities are 80 and 200 $\mu$J/cm$^2$ for GaP and Si.  Traces are offset for clarity.  (c) Frequencies $f$ of reflectivity modulation divided by twice the refractive index, 2$n$, as a function of the probe light wavenumber $\lambda^{-1}$.  The solid lines represent $f/(2n)=v\lambda^{-1}$ with the LA phonon velocity $v$.}
\end{figure}

The interference patterns in the one-color pump-probe measurements [Fig.~\ref{TD400}] decay within several ps for Si and a few tens of ps for GaP.  This is partly because the acoustic pulse moves away from the surface and out of the penetration depth of the probe light [(2$\alpha_\textrm{GaP})^{-1}$=58 nm and (2$\alpha_\textrm{Si})^{-1}$=41 nm].  
To observe the ultrasonic pulses on longer time scales, we probe at longer wavelengths in the visible range, which monitor deeper into the sample, while keeping the pump at 3.1 eV.  Figure~\ref{TDVis}ab compares the oscillatory reflectivity responses from GaP and Si detected at different probe wavelengths $\lambda$.  The traces are modulated with a sinusoidal oscillation, whose frequency $f$ increases with decreasing $\lambda$.  Figure~\ref{TDVis}c plots $f$ divided by twice the refractive index $2n$ as a function of the probe light wavenumber $\lambda^{-1}$.    Linear fits give slopes of $5.84\times10^5$ cm/s for GaP and $8.38\times10^5$ cm/s for Si, which are in good agreement with the LA phonon velocities of GaP and Si [Table~\ref{Const}].  This again confirms that the periodic modulations are caused by the normal strain pulse, as we have already seen on a shorter time scale in Fig.~\ref{TD400}.  

\begin{figure}
\includegraphics[width=0.47\textwidth]{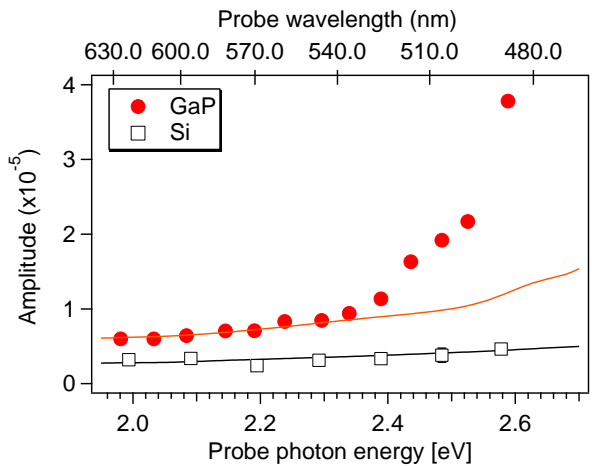}
\caption{\label{Ampl} (Color Online.) Amplitude of the experimental reflectivity oscillation as a function of the probe photon energy $E=\hbar ck$ (symbols).  Pump photon energy is 3.1 eV, and pump densities are 80 and 200 $\mu$J/cm$^2$ for GaP and Si.  Solid curves represent the theoretical amplitudes $A_\textrm{osc}$ scaled to the experimental ones at $E\simeq$ 2.0 eV.}
\end{figure}

The interference patterns are damped on time scales that do not systematically depend on the probe wavelength $\lambda$; $\tau^{-1}$=3.3$\pm$0.5 and 17.6$\pm$2.8 ns$^{-1}$ for GaP and Si when fitted to a damped harmonic oscillation $A_\textrm{osc}\exp(-t/\tau)\sin(2\pi ft+\phi)$.  The observation indicates that the reflectivity oscillation is damped \emph{not} because the acoustic pulses move out of the probed region, since the probing depth $(2\alpha)^{-1}$ varies by a factor of 5 for Si and more than 1000 for GaP over the tuning range of $\lambda$.  We therefore attribute the damping of the oscillations in the two-color measurements to the dephasing caused by the broad bandwidth of the probe light \cite{Maehara14}.  We will confirm this by varying the bandwidth in our theoretical modeling in Sect~\ref{IVB}.

The amplitudes $A_\textrm{osc}$ of the interference patterns, by contrast, depend significantly on $\lambda$ for GaP and moderately for Si, as shown in Fig.~\ref{Ampl}, even though the pump wavelength is kept constant.  The amplitude for GaP is consistently larger than that for Si regardless of $\lambda$. 
The dependence of the oscillation amplitudes on $\lambda$ and on the crystal will be discussed quantitatively in Sect.~\ref{V}.

\section{THEORETICAL MODELING}

\subsection{Generation of Strain Pulse}\label{IVA}

In this section, we present our theoretical modeling of the generation and detection of the strain pulse.
Because the laser spot size (on the order of 10 to100 $\mu$m) is much larger than the penetration depth of the pump light into the semiconductors ($\sim$100 nm), we consider only the one-dimensional distribution of photoexcited carriers along the depth direction.  We also consider only the longitudinal stress and ignore shear stress, because our experiments are on exactly (001)-oriented semiconductors \cite{Matsuda04}.

We first consider the generation of the strain pulse by the deformation potential and the thermoelastic effect, both of which can contribute to the acoustic phonon generation in Si and GaP.  The stress $\Delta \sigma$ induced by the laser pulse is expressed by the sum of the electronic and thermal stresses \cite{Wright01}:
\begin{equation}\label{Th1}
\Delta \sigma(z,t)=-B\dfrac{\partial E_g}{\partial p}N(z,t)-3B\beta\Delta T(z,t),
\end{equation}
where $B$ is the bulk modulous, $E_g$, the band gap, $p$, the pressure, and $\beta$, the linear thermal expansion coefficient.  $N(z,t)$ and $\Delta T(z,t)$ are the photoexcited carrier density and the temperature rise that are dependent on the distance from surface $z$ and time $t$.
The ratio between the electronic and thermal stresses at pump photon energy $\hbar\omega_{pu}$ \cite{Wright01} is approximately given by $(\partial E_g/\partial p)C/[3\beta(\hbar\omega_{pu}-E_g)]\simeq$25 and $14$ for GaP and Si, with the parameters  listed in Table~\ref{Const}. 
We can therefore reasonably neglect the contribution of the thermal stress and consider only the electronic stress via deformation potential coupling with photoexcited carriers in the following calculations.

The elasticity equation that governs the time- and distance-dependences of the stress $\sigma$ is given by:
\begin{equation}\label{Th3}
\rho \dfrac{\partial^2 u(z,t)}{\partial t^2}=\dfrac{\partial \sigma(z,t)}{\partial z}
\end{equation}
with $\rho$ and $u(z,t)$ being the mass density and the lattice displacement.  
The stress consists of an elastic component that is proportional to the strain $\eta\equiv\partial u/\partial z$ and a carrier density-dependent component  \cite{Chern04}:
\begin{equation}\label{Th4}
\sigma (z,t)=\rho v^2 \dfrac{\partial u(z,t)}{\partial z} + a_{cv} N(z,t).
\end{equation}
Here $a_{cv}\equiv-B(\partial E_g/\partial p)$ denotes the relative deformation potential coupling constant, defined by the difference between the coupling constants $a_c$ and $a_v$ of the conduction and valence bands. 

Because the pump pulse is ultrashort (10 to 150 fs) in comparison with the time scale of acoustic phonon propagation, we assume the photoexcitation of carriers to be instantaneous.  
We approximate the photoexcited carrier density by:
\begin{equation}\label{Th5}
N(z,t)= \theta(t)\alpha_\textrm{pu} (1-R_\textrm{pu}) \dfrac{F}{\hbar \omega_\textrm{pu}} e^{-\alpha_\textrm{pu} z}
\end{equation}
with  $\alpha_\textrm{pu}$, $R_\textrm{pu}$ and $F$ being the absorption coefficient, the reflectivity and  the fluence for the pump light.  $\theta$ denotes the step function:
\begin{equation}\label{Th6}
\theta(t)=\begin{cases}
     0 & \text{for}\quad t<0\\
    1 & \text{for}\quad t\geq0
     \end{cases}
\end{equation}
The initial carrier distribution can be modified by diffusion on the picosecond time scale.  
For GaAs, the effect of the ultrafast carrier diffusion on the strain pulse shape is significant \cite{Wright01}, because of the relatively large ambipolar diffusion coefficient, $D_{am}=(\mu_h D_e+\mu_e D_h)/(\mu_e+\mu_h)=12-20$ cm$^2$s$^{-1}$ \cite{Ziebold00,Basak15,Ruzicka10} and of the steep initial distribution of carriers in the depth direction arising from the small optical penetration depth ($\alpha_{pu}^{-1}=$14 nm) of the 3.1-eV pump light.
By contrast, we expect a smaller diffusion coefficient, $D_{am}=5-10$ cm$^2$s$^{-1}$ for both GaP \cite{Ishioka15} and  Si \cite{Sjodin00} in the present excitation conditions.  Moreover, the initial carrier distribution is less steep than that of GaAs because of the larger optical penetration depth ($\alpha\simeq$100 nm).   We therefore neglect the evolution of the carrier distribution by diffusion and its effect on the acoustic pulse shape.

\begin{figure}
\includegraphics[width=0.375\textwidth]{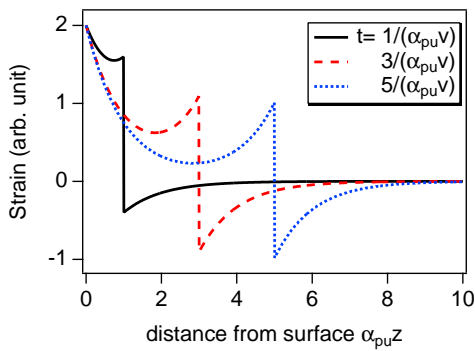}
\caption{\label{Strain} (Color Online.)   Depth profile of the elastic strain calculated with eq. (\ref{Th7}) at different times $t$ after photoexcitation with 3.1-eV optical pulse. The plots are generic for both GaP and Si.}
\end{figure}

We solve the elasticity equation with the initial conditions $u(z,0)$=$\partial u(z,t)/\partial t|_{t=0}$=0 and the boundary condition of zero stress at the surface $\sigma (0,t)=0$.  This yields:
\begin{widetext}
\begin{equation}\label{Th7}
\begin{split}
&u(z,t)= \dfrac{S_0}{2\alpha_\textrm{pu}^2 v^2} \left[\theta(vt-z)\left( e^{-\alpha_\textrm{pu}(z-vt)}-e^{\alpha_\textrm{pu}(z-vt)}\right) -e^{-\alpha_\textrm{pu}(z+vt)}\left( e^{\alpha_\textrm{pu} v t}-1\right)^2\right]\\
&\eta(z,t)=\dfrac{S_0}{2\alpha_\textrm{pu} v^2} \left[ e^{-\alpha_\textrm{pu}(z+vt)}\left( e^{\alpha_\textrm{pu} v t}-1\right)^2-\theta(vt-z)\left( e^{-\alpha_\textrm{pu}(z-vt)}+e^{\alpha_\textrm{pu}(z-vt)}\right)\right]
\end{split}
\end{equation}
\end{widetext}
where $S_0=\alpha_\textrm{pu}^2 a_{cv}(1-R_\textrm{pu})F/(\rho \hbar \omega_\textrm{pu})$.  The time evolution of the calculated strain pulses are shown in Fig.~\ref{Strain}.  In this analysis we do not account for decay process of the strain pulse due to $e.g.$ scattering by photoexcited carriers/phonons and higher-order dispersion of the sound velocity.

We note that, like in InGaN/GaN quantum wells \cite{Sun00} and (111)-oriented GaAs \cite{Vaudel14}, the polar nature of GaP can in principle lead to the acoustic phonon generation through the inverse piezoelectric effect.  The piezoelectric coefficient in the [100] direction is expected to be much smaller than that of the [111] direction, however \cite{BWakata11}.  We therefore disregard the inverse piezoelectric effect as a significant contribution to the generation of the coherent acoustic phonons in GaP.

\subsection{Detection of Strain Pulse}\label{IVB}

Next we calculate the changes in the dielectric constant and the reflectivity due to the photoelastic effect.  We first consider the change of complex dielectric constant $\epsilon$ due to the longitudinal strain pulse $\eta(z,t)$ at probe light wavenumber $k\equiv2\pi\lambda^{-1}$:
\begin{equation}\label{Th8}
\Delta\epsilon(k,z,t)=\frac{\partial\epsilon(k)}{\partial \eta}\eta(z,t).
\end{equation}
Here we ignore $\Delta\epsilon$ due to the surface motion, because it induces only a non-oscillatory change that cannot be separated from the heating and cooling of the electrons and the lattice following the photoexcitation  \footnote{Because the surface displacement is purely real, the surface motion does not contribute to reflectivity change $\Delta R$ anyway.}. The complex reflection coefficient $r_0$ of the semiconductor in the absence of the inhomogeneous strain is expressed by:
\begin{equation}\label{Th9}
r_0(k)=\dfrac{1-\tilde{n}_1(k)}{1+\tilde{n}_1(k)},
\end{equation}
where $\tilde{n}_1$ is the complex refractive index.
The change in the complex reflection coefficient due to the strain is given by \cite{Matsuda02}:
\begin{equation}\label{Th10}
\dfrac{\delta r(k,t)}{r_0}=\dfrac{2ik}{(1-\tilde{n}_1^2)}\int_{0}^{\infty}  dz' \; \dfrac{\partial \epsilon}{\partial \eta} \eta(z',t) e^{2ik_1 z'}
\end{equation}
with $k_1=\tilde{n}_1 k$ being the wavevector of the probe light in semiconductor.  

\begin{figure}
\includegraphics[width=0.48\textwidth]{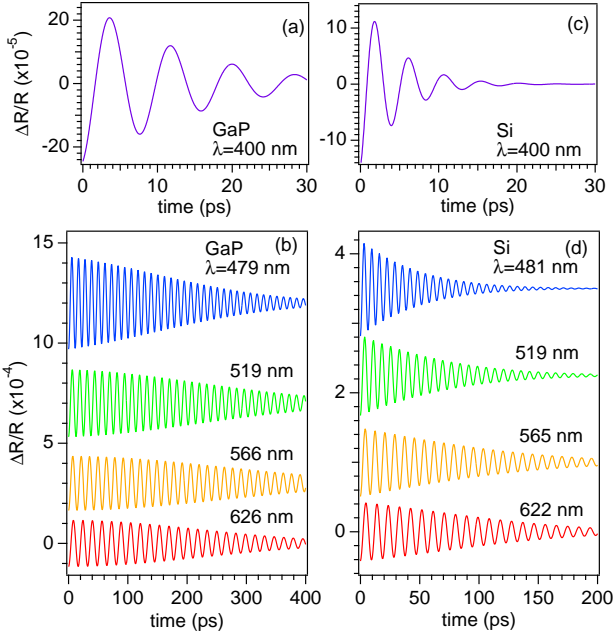}
\caption{\label{Th_Ref} (Color Online.) Theoretical reflectivity changes, 
pumped at 3.1 eV and probed at different wavelength $\lambda$, calculated using eq. (\ref{Th16}): for GaP (a,b) and Si (c,d).  
Traces are offset for clarity. }
\end{figure}

In estimating the modulation of the dielectric constant due to the strain, we assume that it brings about a change in energy gap of the semiconductor and thus modulates the dependence of the dielectric constant on the probe photon energy $E$ \cite{Adachi87,Adachi88}:
\begin{equation}\label{Th11}
\epsilon(E,\eta)\simeq\epsilon(E-a_{cv}\eta).
\end{equation}
Since $a_{cv} \eta$ is significantly smaller than $E$, we can approximate the variation of the dielectric constant with respect to the strain to be:
\begin{equation}\label{Th12}
\frac{\partial\epsilon}{\partial \eta}\simeq-a_{cv}\frac{\partial\epsilon}{\partial E}.
\end{equation}
Eqs.~(\ref{Th11}) and (\ref{Th12}) are good approximations for a direct band gap semiconductor, for which only one ($\Gamma$) satellite valley contributes to $a_{cv}$.  In the present study we extend the approximations to the multi-valley indirect gap semiconductors GaP and Si, though we have no precise knowledge on how $a_{cv}$ depends on $E$, and discuss the validity and limitation of this simple approach in Sect.~\ref{VB}.  We then obtain the modulation of the complex reflection coefficient:
\begin{equation}\label{Th13}
\dfrac{\delta r(k,t)}{r_0}\simeq
-\dfrac{2ik a_{cv}}{(1-\tilde{n}_1^2)}\dfrac{\partial \epsilon}{\partial E}\bigg\vert_{E=\hbar ck}\int_{0}^{\infty}  dz' \; \eta(z',t) e^{2ik_1 z'}.
\end{equation}
The values of $\partial\epsilon/\partial E|_{E=\hbar ck}$ are calculated from the real and imaginary parts of $\epsilon$ in Ref. \cite{Aspnes83}.    Using the expression for the strain pulse [eq. (\ref{Th7})] we get the complex reflectivity change:
\begin{equation}\label{Th14}
\begin{split}
\dfrac{\delta r(k,t)}{r_0}&\simeq
-\dfrac{2ik a_{cv}}{(1-\tilde{n}_1^2)}\dfrac{\partial \epsilon}{\partial E}\bigg\vert_{E=\hbar ck}\dfrac{S_0}{\alpha_\textrm{pu}v^2}   \\
&\times\dfrac{2i\tilde{n}_1 k \left( e^{2i\tilde{n}_1 k v t}-1\right)+\alpha_\textrm{pu}\left(e^{-\alpha_\textrm{pu}vt}-1 \right)}{4 \tilde{n}_{1}^2 k^2+\alpha_\textrm{pu}^2}.
\end{split}
\end{equation}
We can also calculate the probe wavelength-dependence of the oscillation amplitude:
\begin{equation}\label{Th15}
A_\textrm{osc}(k)\simeq\bigg\vert \dfrac{4ik}{(1-\tilde{n}_1^2)}a_{cv}\dfrac{\partial \epsilon}{\partial E}\bigg\vert_{E=\hbar ck}\dfrac{S_0}{\alpha_\textrm{pu}v^2} \dfrac{2i\tilde{n}_1 k  }{4 \tilde{n}_{1}^2 k^2+\alpha_\textrm{pu}^2} \bigg \vert,
\end{equation}

If we calculate the fractional change of the reflected light intensity $\Delta R/R\simeq$2Re$[\delta r/r_0]$ using eq.~(\ref{Th14}), the obtained oscillations [not shown] decay considerably slower than their experimental counterparts. This is because the above expressions are for \emph{monochromatic} probe light, whereas in our experiments the femtosecond probe pulses have finite spectral bandwidths, 0.11 and 0.030$-$0.044 eV (half-width at half-maximum) for one- and two-color pump-probe schemes, respectively.  For such broadband pulses, dephasing among the reflectivity oscillations at different probe wavelengths can no longer be neglected \cite{Maehara14}.  We therefore calculate $\Delta R/R$ by taking into account the  intensity profile $I_{pr}(k)$ of the broadband probe light:
\begin{equation}\label{Th16}
\begin{split}
\dfrac{\Delta R(t)}{R}&=\dfrac{\int dk I_{pr}(k) |r_0+\delta r|^2-\int dk I_{pr}(k) |r_0|^2}{\int dk I_{pr}(k) |r_0|^2}\\
&\simeq\dfrac{\int dk I_{pr}(k) 2\text{Re}[r^*\delta r]}{\int dk I_{pr}(k) |r_0|^2}.
\end{split}
\end{equation}
Fig.~\ref{Th_Ref} plots the oscillatory part of $\Delta R/R$ for GaP and Si pumped at 400-nm and probed at different wavelengths, calculated with eq. (\ref{Th16}).  Here we use the relative deformation potential coupling constants $a_{cv}$ corresponding to the interband transitions $\Gamma^v\rightarrow\Gamma^c$ for GaP and $L^v\rightarrow L^c$ for Si, whose values are listed in Table~\ref{DP}, for both the pump and probe interactions.   The calculations reproduce the damping and the frequency of the experimental reflectivity oscillations for both GaP and Si in Figs. \ref{TD400} and \ref{TDVis} quite well.
We note that the net effect of eq.~(\ref{Th16}) can be reduced to multiplying eq.~(\ref{Th14}) by $\exp[-(n v \Delta kt)^2/\ln(2)]$, if $n$ and $\partial\epsilon/\partial E$ vary slowly within the full width $2\Delta k$ of the probe pulse spectrum.   This additional term makes the reflectivity oscillation be damped out faster than the exponential damping due to the strain packet leaving the probed region.  The additional damping is faster for GaP  than for Si at a given probe spectrum, because both its $n$ and $v$ are smaller.

Our theoretical modeling fails to reproduce the \emph{amplitude} of the reflectivity oscillation for GaP, which in the experiment [Fig.~\ref{TDVis}a] is significantly larger at $\lambda$=479 nm than at other $\lambda$.  The discrepancy is also seen in the $A_\textrm{osc}$ vs $E$ plot in Fig.~\ref{Ampl}, where the experimental amplitude of GaP increases much more rapidly with $E$ than the calculation.
Because the $E$-dependences of other parameters in eqs.~(\ref{Th14}) and (\ref{Th15}) are known, the discrepancy suggests that our choice of $a_{cv}$ needs to be reconsidered for GaP, as we will discuss in Sect.~\ref{VB}.

\begin{table}
\caption{\label{DP} Relative deformation potential coupling constant $a_{cv}$ in unit of eV for major optical transitions taken from literature.}
\begin{ruledtabular}
\begin{tabular}{lcccccc}
&$\Gamma^v\rightarrow \Gamma^c$ &$\Gamma^v\rightarrow X^c$&$\Gamma^v\rightarrow L^c$&$L^v\rightarrow L^c$ &$L^v\rightarrow X^c$&Ref.\\
\hline
Si	&-13.02& 2.22&&-4.52&-1.24&\cite{Cardona87}\\
	&-0.48& 1.72&&&&\cite{vdWalle89}\vspace{4mm}
\\
GaP	&-7.83& 2.25&-3.07&&&\cite{Cardona87}\\
	&-8.83& 1.56&&&&\cite{vdWalle89}\\
\end{tabular}
\end{ruledtabular}
\end{table}

\section{Discussion}

\subsection{Delayed build-up of acoustic pulse}\label{VA}
Although our theoretical modeling is simple, it can reveal some basic physics behind the generation and the detection of the coherent acoustic phonons.  

\begin{figure}
\includegraphics[width=0.45\textwidth]{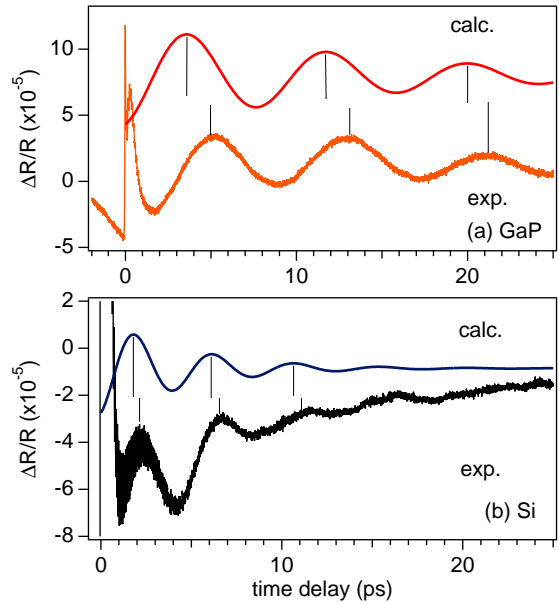} 
\caption{\label{ET}  (Color Online.) Experimental and calculated reflectivity changes for GaP (a) and Si (b) pumped and probed with 10-fs pulses at 3.1 eV.  Experimental pump densities are 30 and 90 $\mu$J/cm$^2$ for GaP and Si.  The calculated reflectivity traces are scaled to the experimental ones 
and offset for clarity.  Vertical lines show the positions for the oscillation maxima.}
\end{figure}

Figure~\ref{ET} compares the experimental and calculated $\Delta R/R$ for GaP and Si pumped and probed with 3.1-eV, 10-fs pulses.  
We see that the extrema of the experimental $\Delta R/R$ come later than the theoretically calculated ones, by $\sim$1.2 ps for GaP and by $\sim$0.4 ps for Si.  Our two-color experiments with tunable visible probe light [Fig.~\ref{TDVis}] also obtain delays with respect to the calculations [Fig.~\ref{Th_Ref}b,d], by 1.4$\pm$0.2 ps for GaP and 0.3$\pm$0.2 ps for Si, that are not systematically dependent on the probe photon energy $E$.  The insensitivity of the phase delay to $E$ and its dependence on the crystal confirm that it is not an aspect of the detection process but actually arises from a pump-induced dynamics.    We therefore consider the observed phase delay to be caused by the delayed build-up of the strain pulse in both GaP and Si.  

There are several possible origins for the build-up time for the strain pulse on the picosecond time scale, including i) thermoelastic effect, ii) inverse piezoelectric effect, and iii) intervalley carrier scattering.  As we have already discussed in Sect.~\ref{IVA}, the thermoelastic effect in the acoustic pulse generation in GaP and Si  is an order of magnitude smaller than the deformation potential coupling.  Moreover, the creation of acoustic phonons via thermoelastic effect takes longer; the acoustic phonon temperature rises in $\lesssim$10 ps after photoexcitation in the $L$ valley of Si \cite{Gunnella16} and presumably on a similar time scale for GaP \cite{Othonos98}. 
The inverse piezoelectric effect \cite{Ruello15,Sun00} for GaP, by contrast, would be significantly faster than the experimentally observed phase delay, since the screening of the pre-existing field takes $\sim$0.1 ps \cite{Ishioka15}; in addition, the mechanism does not exist in the centrosymmetric Si.  We therefore neglect the thermoelastic and inverse piezoelectric effects as the possible origins of the experimentally observed phase delay.

In our theoretical modeling we have assumed that the lattice deforms instantaneously by coupling with photoexcited carriers through a uniquely defined deformation potential coupling constant $a_{cv}$ [eq.~(\ref{Th7})].  This is a reasonable approximation in direct band gap semiconductors, in which electrons photoexcited in the $\Gamma$ valley mostly remain there until they recombine with holes.  
In indirect gap semiconductors, however, the situation is more complicated because most of the photoexcited electrons undergo intervalley scattering and populate different satellite valleys at different times.
For Si, previous studies \cite{Ichibayashi11,Sangalli15} revealed that the electrons excited at the $L$ valley are scattered into the $X$ valley, as illustrated in Fig.~\ref{band}b,  with a time constant of 0.2$-$0.5 ps.  
For GaP, the electrons excited at the $\Gamma$ valley are scattered almost instantaneously ($\lesssim$0.1 ps) into the lower $X_1$ and $L$ satellite valleys \cite{Cavicchia95,Collier13,Sjakste07}, as illustrated in Fig.~\ref{band}a.  The electrons have additional, slower relaxation processes with time constants of $\sim$0.7 and 4 ps, which the previous studies attributed to the intervalley scattering from $\Gamma$ into the upper $X_3$ valley and the relaxation from the $X_3$ to the $X_1$ \cite{Collier13,Sjakste07}. These longer time scale processes might also have contributions from the $L$ to $X$ intervalley scattering.   In both semiconductors, the build-up times of the strain observed in the present study (0.4 and 1.2 ps for Si and GaP) falls within the time scale of the population build-up at the conduction band minima (CBM) in the $X$ valleys.  The agreement suggests that a significant contribution to the strain generation results from coupling with electrons at the CBM because of their long decay times compared with the time scale of lattice deformation.  
To fully discuss the ultrafast dynamics of the strain build up, however, we need to develop a theoretical model that takes  into account the valley-dependent deformation potential coupling, which is beyond the scope of the present study.

\subsection{Deformation potentials in acoustic pulse generation and detection}\label{VB}

The amplitudes of the experimental reflectivity oscillations are consistently larger for GaP than for Si, with the ratio $A_\textrm{osc}^\textrm{GaP}/A_\textrm{osc}^\textrm{Si}$=4.9 for the one-color pump-probe measurements at 3.1 eV after we normalize the signals to account for the differences in pump intensity.  From this value we can estimate the ratio of the deformation potential couplings of the two semiconductors by using eq.~(\ref{Th15}), which includes $a_{cv}$ explicitly for the detection as well as through $S_0$ for the generation of the acoustic pulse.  By assuming the same value of $a_{cv}$ for the generation and the detection for the one-color experiments, we obtain $|a_{cv}^\textrm{GaP}/a_{cv}^\textrm{Si}|\simeq$2.4.  

The obtained ratio can be compared with theoretical predictions   \cite{Cardona87,vdWalle89}.
For GaP and Si, the 3.1-eV pump light causes the $\Gamma^v\rightarrow\Gamma^c$ and $L^v\rightarrow L^c$ interband transitions, respectively, as illustrated by the vertical arrows in Fig.~\ref{band}.  The ratio of $a_{cv}$ corresponding to these transitions is $|a_{cv}^\textrm{GaP}/a_{cv}^\textrm{Si}|$=1.7, which agrees reasonably with the experimental estimation, 2.4.   In the both semiconductors, the electrons are scattered to the $X$ (and $L$) valleys, as shown with broken arrows in Fig.~\ref{band}.  The ratio between the indirect transitions, $\Gamma^v\rightarrow X^c$ for GaP and $L^v\rightarrow X^c$ for Si, $|a_{cv}^\textrm{GaP}/a_{cv}^\textrm{Si}|$=1.8, is still in reasonable agreement with the experiment.  We note that the theoretical ratio is far from the experimental one if we consider the transitions from the $\Gamma$ valley of Si: $|a_{cv}^\textrm{GaP}/a_{cv}^\textrm{Si}|$=0.6 (18) for $\Gamma^v\rightarrow\Gamma^c$ and 1.0 (0.9) for $\Gamma^v\rightarrow X^c$according to Ref.~\cite{Cardona87} (\cite{vdWalle89}).  
 
%\subsection{Deformation potential in detection}\label{VC}

Our theoretical modeling reproduces well the experimentally observed variation of the $A_\textrm{osc}$ as a function $E$ for Si over the whole range of $E$ in Fig.~\ref{Ampl}.  For GaP, by contrast, the calculation reproduces the experiment only for $E\lesssim$2.3 eV, if we scale the theoretical and experimental amplitudes at the lowest $E$.  Above 2.3 eV, the experimental amplitude increases more rapidly with $E$ than the theoretical expectation, and is three times larger than the theoretical one at $E=$2.6 eV, as shown in Fig.~\ref{Ampl}.

The discrepancy can be attributed to our neglect of the $E$-dependence of $a_{cv}$ in eq.~(\ref{Th11}), which is reasonable for direct semiconductors but less reasonable for indirect gap semiconductors, especially when $E$ is varied around the band gap energy. 
For Si we expect $a_{cv}$ to be nearly independent of $E$ in the range of 2.0$-$2.6 eV,  which is far from both indirect and direct band gap energies. 
For GaP, by contrast, the range of $E$ for the two-color measurements includes the indirect gap energy $E_g$=2.26 eV and close to the direct gap energy $E_{dir}=$2.78 eV at room temperature  \cite{Lorenz68,Panish69,Humphreys78,Takizawa83}.  
We therefore expect multiple optical transitions to be involved in the detection in this energy range, with their relative contributions depending on $E$, which leads to significant $E$-dependence of $a_{cv}$.  The experimentally observed $E$-dependence can be semi-quantitatively explained if we consider the contribution of the indirect ($\Gamma\rightarrow\ X^c$) and the direct ($\Gamma\rightarrow\Gamma^c$) transitions to dominate the detection below and above $E_g$.  $|a_{cv}|$ of the latter is 3.5 times larger than the former, which is in good agreement with the three times discrepancy between the experiment and theory at the highest $E$. 
To obtain the precise $E$-dependence of $a_{cv}$, however, is beyond the scope of the present study.  

\section{Conclusion}
 
We have studied the picosecond dynamics of coherent acoustic phonons generated at the (001)-oriented GaP and Si surfaces in pump-probe reflectivity scheme.  The generation of a normal strain pulse has been theoretically modeled in terms of the deformation potential coupling, and its detection in terms of the photoelastic effect.  
Comparison between the experiment and the theory has revealed that the oscillation damping occurs through dephasing between different frequency components of the probe light before the acoustic pulse moves out of the optical probing depth.
The deformation potential couplings in GaP has been estimated to be twice as strong as in Si based on the experimental amplitudes of the reflectivity oscillations.  
The shift in the dielectric constant under stain is nearly independent of photon energy for Si, whereas its energy-dependence cannot be ignored for GaP, when the photon energy is swept between 2.0 and 2.6 eV.
The strain pulse builds-up over finite time for both GaP and Si that is comparable with the time for intervalley scattering into the lowest conduction valley in each semiconductor.
Our study thus demonstrates the general applicability of coherent acoustic phonon spectroscopy to polar and non-polar semiconductors via interband excitation without the recourse to additional phonon transduction structures.
This makes our method broadly applicable to ultrafast opto-acoustical measurements of fundamental and practical nature.

\begin{acknowledgments}
This work is partly supported by NSF grants DMR-1311845 (Petek) and DMR-1311849 (Stanton), as well as by the Deutsche Forschungsgemeinschaft through SFB 1083 and HO2295/8.  
\end{acknowledgments}

\bibliographystyle{apsrev4-1}
\bibliography{AcousticBulk}

\end{document}